\documentclass[11pt]{article}
\usepackage{amsfonts}
\usepackage{amssymb}
\newcommand{\be}{\begin{equation}}
\newcommand{\ee}{\end{equation}\noindent}
\newcommand{\bear}{\begin{eqnarray}}
\newcommand{\ear}{\end{eqnarray}\noindent}
\newcommand{\no}{\noindent}
\date{}

\newcommand{\slD}{\raise.15ex\hbox{$/$}\kern-.57em\hbox{$D$}}
\newcommand{\slpartial}{\raise.15ex\hbox{$/$}\kern-.57em\hbox{$\partial$}}
\newcommand{\slG}{{{\dot G}\!\!\!\! \raise.15ex\hbox {/}}}

\def\GBd12{{\dot G}_{B12}}

\def\non{\nonumber}
\def\beqn*{\begin{eqnarray*}}
\def\eqn*{\end{eqnarray*}}

\def\square{\kern1pt\vbox{\hrule height 1.2pt\hbox{\vrule width 1.2pt
   \hskip 3pt\vbox{\vskip 6pt}\hskip 3pt\vrule width 0.6pt}
   \hrule height 0.6pt}\kern1pt}

\def\slash#1{#1\!\!\!\raise.15ex\hbox {/}}

\def\dps{\displaystyle}

\def\half{{1\over 2}}

\def\4piTD{{(4\pi T)}^{-{D\over 2}}}
\def\4piT4{{(4\pi T)}^{-2}}

\def\Tintm4{{\dps\int_{0}^{\infty}}{dT\over T}\,e^{-m^2T}
    {(4\pi T)}^{-2}}
\def\Tintm{{\dps\int_{0}^{\infty}}{dT\over T}\,e^{-m^2T}}

\def\bbbz{{\mathchoice {\hbox{$\sf\textstyle Z\kern-0.4em Z$}}
{\hbox{$\sf\textstyle Z\kern-0.4em Z$}}
{\hbox{$\sf\scriptstyle Z\kern-0.3em Z$}}
{\hbox{$\sf\scriptscriptstyle Z\kern-0.2em Z$}}}}
\newcommand{\hoch}[1]{$\, ^{#1}$}

\newcommand{\auth}{\large B. Eden\hoch{a}, A. C. Petkou\hoch{b},
C. Schubert\hoch{a,c}
\footnote{Address after 1. 10. 2000:
Instituto de Fisica y Matematicas,
Universidad Michoacana de San Nicolas de Hidalgo,
Morelia, Mexico.}
, E. Sokatchev\hoch{a}}
\begin{document}
\thispagestyle{empty}

\hfill{LAPTH-811/2000, KL-TH 00/06}

\vspace{20pt}

\begin{center}
{\Large{\bf Partial non-renormalisation of the stress-tensor
four-point function in $N=4$ SYM and AdS/CFT}}
\vspace{30pt}

\auth

\vspace{15pt}

\begin{itemize}

\item [$^a$] {\it\small Laboratoire d'Annecy-le-Vieux de Physique
Th{\'e}orique\footnote{UMR 5108 associ{\'e}e {\`a}
 l'Universit{\'e} de Savoie} LAPTH, Chemin de Bellevue, B.P. 110,
F-74941 Annecy-le-Vieux, France}
\item [$^b$] {\it \small Department of Physics, Theoretical Physics,
    University of Kaiserslautern, Postfach 3049, 67663 Kaiserslautern,
    Germany}
\item [$^c$] {\it \small California Institute for Physics and
Astrophysics,
366 Cambridge Avenue, Palo Alto, CA 94306, US}
\end{itemize}

\vspace{60pt}

{\bf Abstract}

\end{center}
We show that, although the correlator of four stress-tensor
multiplets in $N=4$ SYM is known to have radiative corrections,
certain linear combinations of its components are protected from
perturbative renormalisation and remain at their free-field
values. This result is valid for weak as well as for strong
coupling and for any gauge group. Our argument uses Intriligator's
insertion formula, and includes a proof that the possible contact
term
contributions cannot change the form of the amplitudes. \\
Combining this new non-renormalisation theorem with Maldacena's
conjecture allows us to make a prediction for the structure of
the corresponding correlator in AdS supergravity. This is
verified by first considerably simplifying the strong coupling
expression obtained by recent supergravity calculations, and then
showing that it does indeed exhibit the expected structure.

{\vfill\leftline{}\vfill

\pagebreak

\section{Introduction}

Much effort has been devoted over the years to the study of the
dynamical aspects of quantum field theory. Weak coupling
expansions have been pushed to high orders providing useful
insight, however, the strong coupling regime of most theories has
remained elusive. Nevertheless, important information can be and
has been obtained from the study of various quantities whose weak
and strong coupling behaviour is accessible. Such quantities are
axial \cite{axial} and conformal anomalies
\cite{henske,bafrts},
and also coupling constants that are not renormalised as one goes
from weak to strong coupling \cite{lmrs,dfs,petske}. The
emergence of the latter quantities has been particularly noted in
the context of the AdS/CFT correspondence conjectured by
Maldacena \cite{adscft1, adscft2, adscft3}.
One of the most exciting features of
this conjecture is that strong coupling information for certain
quantum field theories can be obtained from tree-level
supergravity calculations. Therefore, with the explicit strong coupling
results for various correlation functions in hand, one
can identify ``non-renormalised" quantities which remain the same
both in the weak and the strong coupling regimes.

So far the strongest evidence in favour of this conjecture has
been gathered in the context of $N =4$ SYM theory with gauge group
$SU(N_c)$ \cite{fmmr,cnss,lmrs,liutse,dfs,ken1,dfmmr1}. This is
not surprising since the study of this theory has a long history
and its dynamics is relatively well understood. Moreover, the
success of this programme has prompted more thorough
investigations of $N=4$ SYM within field theory, and in particular
a search for non-renormalisation theorems for the correlators
involving only ``short" operators, i.e. certain class of gauge
invariant composite operators which do not depend on all the
Grassmann variables in superspace. The most typical example,
first introduced in \cite{hw1}, are the series of operators
obtained by tensoring the $N=4$ SYM field strength considered as
a Grassmann analytic harmonic superfield. They were identified
with short multiplets of $SU(2,2/4)$  and their correspondence
with the K-K spectrum of IIB supergravity was established in
\cite{andfer}. Short multiplets are important in the $AdS/CFT$
correspondence because they have protected conformal dimensions
and therefore allow a reliable comparison between quantities
computed in the bulk versus quantities derived in the $CFT$
\cite{fpz}.

Recently it has been found that, for correlators of short
multiplets, the absence of radiative corrections is a rather
common phenomenon; by now non-renormalisation theorems have been
established not only for two- and three-point functions
\cite{dfs,hsw,ken1,gkp,intski,ehw,skiba,psz}, but also for
so-called ``extremal" \cite{dfmmr2,biakov,ehssw3} and
``next-to-extremal" \cite{ehssw3,erdper,ehsw} correlators; these
are $n$-point functions obeying certain conditions on the
conformal dimensions of the operators involved.

At the same time, some correlators of short operators are known to
acquire quantum corrections beyond tree level. The simplest
example is the four-point function of $N=4$ SYM supercurrents, as
has been shown by explicit computations at two loops in
\cite{gps,ehssw1,ehssw2}, and at three loops in \cite{ess,bkrs}
\footnote{In our terminology the free-field and one-loop
contributions to this correlator are synonymous, the order
$O(g^2)$ contribution corresponds to two loops, etc.}.
Nevertheless, even in this case one can still formulate a
``partial non-renormalisation" theorem which is the main subject
of this paper.

%\cite{ehssw1,ehssw2,hssw,EHPSW,ess}

The work presented here is part of an ongoing investigation of
this four-point function, which grew out of a line of work
initiated by P.S. Howe and P.C. West. Their original aim was to
study the implications of superconformal covariance for
correlators satisfying Grassmann and harmonic analyticity
constraints. In references \cite{hw1,hw2} a systematic
investigation of the superconformal Ward identities and their
consequences for Greens functions of $N=2$ and $N=4$ short
operators was initiated \footnote{Similar works analysing
constrained $N=1$ superfields are
\cite{cowe,osborn,piwe,kuzthe,dolosb}.}. The $N=2$ operators are
gauge-invariant products of the hypermultiplet and the $N=4$
operators are gauge-invariant products of the $N=4$ field
strength. A number of interesting results were derived
\cite{hw1,hw2}, in particular: The $N=4$ SYM field strength is a
covariantly analytic scalar superfield from which the
aforementioned set of analytic gauge-invariant operators can be
built; the two- and three-point Greens functions of these
operators were determined up to constants \cite{hw1,hsw} (see also
\cite{lmrs,dfs,gkp,intski}); the set of all non-nilpotent
analytic superconformal invariants was given \cite{hw2}. Finally,
in \cite{EHPSW} it was shown how to derive certain differential
constraints on the correlator of four stress-tensor multiplets
using only the superconformal algebra and some general properties
of the $N=2$ \cite{hh} or $N=4$ \cite{hhh} harmonic superspace
formulations of $N=4$ SYM. (These constraints are reproduced and
their general solution is obtained in the Appendix to the present
paper.)

However, we will see that considerably stronger restrictions can
be derived for this amplitude if not only pure symmetry-based
arguments are employed, but also some direct input from field
theory making use of the explicit form of the $N=4$ SYM
Lagrangian written down in terms of $N=2$ superfields. The
essential tool in our analysis is Intriligator's ``insertion" (or
``reduction") formula, which relates the above four-point
function to a five-point function involving the original
operators and the gauge-invariant $N=2$ SYM Lagrangian. This
procedure, introduced in the present context in \cite{ken1}, has
turned out both more efficient and more powerful than the direct
approach to four-point functions: The lowest term of the
five-point function obtained in this way in the context of $N=2$
harmonic superspace is a nilpotent superconformal covariant of
mixed chiral-analytic type. Such objects are more strongly
constrained than the original non-nilpotent amplitudes. The
insertion approach has been constructively used in the harmonic
superspace derivation or rederivation of some of the
non-renormalisation theorems mentioned before
\cite{ehw,ehssw3,ehsw}.

In the past the applicability of Intriligator's insertion formula
has been questioned on the grounds that possible contact term
contributions might spoil any prediction based on it
\cite{petske,hssw}. While contact term contributions to
correlators can usually be consistently ignored, it had been
pointed out in \cite{petske} that this is not obvious if the
insertion procedure is used: It involves an integration which may
promote a contact term to a regular term. Here we
resolve this issue by finding necessary conditions for the
existence of a five-point contact term with the required
properties and by giving its most general allowed form. It then
becomes clear that the regular term which it produces upon
integration is compatible with our non-renormalisation statement.
%Note, however, that
%this term, while apparently absent up to the three-loop level,
%could indeed potentially spoil the usefulness of the insertion
%procedure at even higher loop orders.

The insertion formula has also proved very useful in explicit
quantum calculations at two and three loops because it allows to
apply superconformal covariance arguments to significantly reduce
the complexity of Feynman diagram calculations of the correlator
discussed here in $N=2$ harmonic superspace \cite{hssw,ess}. The
results obtained show a remarkable pattern: certain linear
combinations of the amplitudes are protected from perturbative
renormalisation, and thus remain at their free-field values. As
we show in the present paper, this property can be generalised to
every loop order and thus to a non-perturbative result,
irrespectively of the choice of the gauge group. Combining this
new non-renormalisation theorem with Maldacena's conjecture we can
make a prediction for the structure of the strong coupling limit
obtained from AdS supergravity.

AdS supergravity calculations have been initiated in
\cite{fmmr,lmrs,liutse,dhofre,chasch,dfmmr1,dfmmr2} and most of the
relevant methodology was developed in these articles. Yet, there
the focus was on
correlators whose CFT counterparts involve the top components
$F^2$, $F\tilde F$ of the $N=4$ stress-tensor multiplet, since
these are more readily accessible from the supergravity side. On
the contrary, on the CFT side it is easiest to investigate the
lowest components (the physical scalars) of the same multiplet.
So, until recently the results on both sides were difficult to
compare. With the completion of the computation of the quartic
terms in the supergravity effective action, a strong coupling
limit for the lowest component of the SYM stress-energy tensor
four-point function became available \cite{AF}. However, it is
presented in a form which still contains parameter integrals.
To verify our non-renormalisation prediction,
we first bring this result into a completely explicit
form, involving only logarithms and dilogarithms of the conformal
cross ratios. The ``non-renormalised part" is then indeed found
to agree with the free-field SYM amplitude.

The organisation of the paper is as follows: In Section 2,
exploiting $SO(6)$ and conformal covariance and point-permutation
symmetry we show how to reduce the lowest component of the
correlator of four stress tensors in $N=4$ SYM to that of a
single $N=2$ hypermultiplet correlator. Section 3 provides a
minimum of information about the $N=2$ harmonic superspace
formulation \cite{hh} of $N=4$ SYM. Section 4 is central, and
contains the proof of our ``partial non-renormalisation theorem"
based on the insertion formula. This section also includes the
necessary study of the possible contact terms. In Section 5 we
verify that the AdS/CFT correspondence holds for the
non-renormalised part of the correlator. In the Appendix we state
the differential constraints on this four-point function found by
the more abstract analysis of \cite{EHPSW}. We give their general
solution, from which it is evident that these constraints are
weaker than the ones obtained in Section 4.

\setcounter{equation}0 \section{The $N=4$ SYM four-point
stress-tensor correlator}\label{N4N2}

Here we show how one can compute the leading scalar term of the
$N=4$ four-point function of four SYM supercurrents (stress-tensor
multiplets) from the leading scalar term of an $N=2$
hypermultiplet four-point function. This section is based on
\cite{ehssw1} but we also derive some additional restrictions on
the amplitude following from point-permutation symmetry.

We recall that in $N=4$ superspace the $N=4$ field strength
superfield $W^A,\ A=1,\ldots 6$ transforms under the vector
representation of the R symmetry group $SO(6)\sim SU(4)$. This
superfield satisfies an on-shell constraint reducing it to six
real scalars, four Majorana spinors and a vector.

The $N=4$ supercurrent is given by
\begin{equation}
 T^{AB}=W^A W^B -{1\over 6}\delta^{AB}W^C W^C
\end{equation}
(the trace over the Yang-Mills indices is implied). It is in the
symmetric traceless ${\bf 20}$ of $SO(6)$ and is conserved as a
consequence of the on-shell constraints on $W^A$.

The four-point function we are going to consider is
\begin{equation} G^{(N=4)}=\langle
T^{A_1B_1}T^{A_2B_2}T^{A_3B_3}T^{A_4B_4}\rangle\label{N4corre}
\end{equation}
where the numerical subscripts indicate the point concerned. This
function can be expressed in terms of $SO(6)$ invariant tensors
multiplied by scalar factors which are functions of the
coordinates. Given the symmetry of $G^{(N=4)}$, the only $SO(6)$
invariant tensor that can arise is the Kronecker $\delta$, and
there are two modes of hooking the indices up, each of which can
occur in three combinations making six independent amplitudes in
all. \footnote{An alternative explanation is provided by a double
OPE. Indeed, each pair of operators is decomposed into six irreps
of $SO(6)$, ${\bf 20}\times {\bf 20} = {\bf 1} + {\bf 15} + {\bf
20} + {\bf 84} + {\bf 105} + {\bf 175}$. Then the double OPE,
being diagonal, gives rise to six structures.} Thus, for the
leading component in the $\theta$ expansion \footnote{In section
\ref{HM} we shall show that superconformal covariance allows us
to restore the complete $\theta$ dependence of (\ref{N4corre}),
given its leading component.} we have
\begin{eqnarray}
G^{(N=4)}|_{\theta=0}&=&a_1(s,t){(\delta_{12})^2(\delta_{34})^2
\over x^4_{12}x^4_{34}} + a_2(s,t){(\delta_{13})^2(\delta_{24})^2
\over x^4_{13}x^4_{24}}
 +a_3(s,t){(\delta_{14})^2(\delta_{23})^2 \over
x^4_{14}x^4_{23}} \label{sixcoeff}\\
 &\phantom{=}&
 +b_1(s,t){\delta_{13}\delta_{14}\delta_{23}\delta_{24} \over
x^2_{13}x^2_{14}x^2_{23}x^2_{24}}+
 b_2(s,t){\delta_{12}\delta_{14}\delta_{32}\delta_{34} \over
x^2_{12}x^2_{14}x^2_{32}x^2_{34}}
 +b_3(s,t){\delta_{12}\delta_{13}\delta_{42}\delta_{43} \over
x^2_{12}x^2_{13}x^2_{42}x^2_{43}} \nonumber
\end{eqnarray}
where $x^2_{pq}= (x_p-x_q)^2$ and, for example,
\begin{equation}
(\delta_{12})^2(\delta_{34})^2=\delta_{\{A_1B_1\}}^{A_2B_2}
\delta_{\{A_3B_3\}}^{A_4B_4}\;, \qquad
\delta_{13}\delta_{14}\delta_{23}\delta_{24}=\delta_{\{A_1B_1\}}^{\{A_3\{B_4}
\delta_{\{A_2B_2\}}^{A_4\}B_3\}}\;,
\end{equation}
and where the braces denote tracefree symmetrisation at each
point.

When writing down (\ref{sixcoeff}) we have taken into account the
conformal covariance of the correlator. In each term we have
introduced the corresponding scalar propagator structure which has
the required conformal weight of the correlator. This implies that
the coefficient functions $a_{1,2,3}, \ b_{1,2,3}$ are conformal
invariants, so they depend on the two conformal cross-ratios
\begin{equation}\label{crrt}
 s = {x^2_{12}x^2_{34}\over x^2_{13}x^2_{24}}\;, \qquad
t = {x^2_{14}x^2_{23}\over x^2_{13}x^2_{24}}\;.
\end{equation}

Next, we notice the point-permutation symmetry of the correlator
(\ref{N4corre}). To implement it it is sufficient to require
invariance under two permutations, for instance,
\begin{equation}\label{permut13}
  1 \  \rightarrow \ 3\ : \ s \  \rightarrow \ t
  \qquad \mbox{and}\qquad 1 \  \rightarrow \ 2\ :
  \ s \  \rightarrow \  {s\over t}\ , \ t \  \rightarrow
  \ {1\over t}\ .
\end{equation}
This leads to the following constraints:
\begin{eqnarray}
  &&a_1(s,t) =  a_3(t,s) =  a_1(s/t,1/t)\nonumber\\
  &&a_2(s,t) = a_2(t,s) = a_3(s/t,1/t)\nonumber\\
  &&b_1(s,t) = b_3(t,s) =  b_1(s/t,1/t)\nonumber\\
  &&b_2(s,t) = b_2(t,s) =  b_3(s/t,1/t)\label{permut12}
\end{eqnarray}

So, the six coefficients in the $N=4$ amplitude are in fact
reduced to only {\it two independent} ones, one of the $a_i$ and
one of the $b_i$. Now we shall show that one can determine these
two functions by studying a certain $N=2$ component of the $N=4$
correlator. Let us see what happens when one reduces $N=4$
supersymmetry to $N=2$. The first step is to decompose the
$SO(6)$ vector $W^A$ in a complex basis as $3+\overline 3$ under
$SU(3)$ and further as $2+1+\overline 2 + \overline 1$ under
$SU(2)$:
\begin{equation}W^A \ \rightarrow\ (W^i\equiv \phi^i, W^3\equiv W,
W_i \equiv \bar \phi_i, W_3 \equiv \overline W)\;.
\end{equation}
Here $W$ is the $N=2$ SYM field-strength and $\phi^i$ ($i=1,2$) is
the $N=2$ matter hypermultiplet.

{}From the on-shell constraints on $W^A$ it is easy to derive that
these superfields (evaluated at $\theta_3=\theta_4=0$) obey the
corresponding constraints,
\begin{eqnarray}
  && \bar D_{\dot\alpha\; i}\; W = 0\;, \label{consW1}\\
  && \epsilon^{\alpha\beta}\; D_{\alpha}^i D^{ j}_\beta
W=0 \label{consW2}
\end{eqnarray}
and
\begin{equation}\label{hmdef}
  D_{\alpha}^{(i}\; \phi^{j)}=0\;, \qquad
  \bar D_{\dot\alpha\; i}\; \phi^j - {1\over 2}\delta^j_i\;
  \bar D_{\dot\alpha\; k}\; \phi^k =0\;.
\end{equation}
Eq. (\ref{consW1}) means that $W$ is a chiral superfield (a
kinematic constraint) while eq. (\ref{consW2}) is the Yang-Mills
equation of motion, and eqs. (\ref{hmdef}) are the field equations
of the hypermultiplet. \footnote{Strictly speaking, eqs.
(\ref{consW1})-(\ref{hmdef}) hold only for an Abelian gauge
theory and in the non-Abelian case a gauge connection needs to be
included. However, the constraints satisfied by the
gauge-invariant bilinears that we shall consider are in fact the
same in the Abelian and non-Abelian cases.}

In $N=2$ harmonic superspace (see section \ref{HSS} for a review)
all $SU(2)$ indices are projected out by harmonic variables thus
obtaining objects which carry $U(1)$ charge but are singlets under
$SU(2)$. For instance, the harmonic superfield $q^+$ and its
conjugate $\widetilde q^+$ are related to the ordinary
hypermultiplet $N=2$ superfield $\phi^i$ by
\begin{equation}\label{trivsol}
 q^+= u^+_i\phi^i\;, \qquad \widetilde q^+= u^{+i}\bar\phi_i
\end{equation}
on shell.

For our purpose of identifying the coefficients in the $N=4$
amplitude it will be sufficient to restrict $T^{AB}$ to the
following $N=2$ projections involving only hypermultiplets:
\begin{equation}
 T_{ij} = \bar\phi_i \bar\phi_j \ (\mbox{ at
points 1 and 3})\;,  \qquad  T^{ij}=\phi^i\phi^j \
(\mbox{ at points 2 and 4})\;.
\end{equation}
Then we multiply the resulting hypermultiplet correlators by
$u^+_i u^+_j$ at each point and obtain
\begin{equation}\label{1st1}
  G^{(N=2)} = \langle \widetilde q^+\widetilde
q^+|q^+q^+|\widetilde q^+ \widetilde q^+ |q^+q^+\rangle\;.
\end{equation}
Its leading component is
\begin{equation}\label{1st}
  G^{(N=2)}|_{\theta=0} =  a_1(s,t){(12)^2(34)^2\over x^4_{12}x^4_{34}}  +
a_3(s,t){(14)^2(23)^2\over x^4_{14}x^4_{23}}  +
b_2(s,t){(12)(23)(34)(41)\over x^2_{12}x^2_{23}x^2_{34}x^2_{41}}
\end{equation}
where, for example,
$$
(12)= u_1^{+i}u_2^{+j}\epsilon_{ij}\;.
$$

It is now clear that if we know the coefficient functions in the
$N=2$ correlator (\ref{1st}), using the symmetry properties
(\ref{permut12}) we can obtain all the six coefficients in the
$N=4$ amplitude (\ref{sixcoeff}). Recalling that the symmetry of
the correlator under the exchange of points $1\leftrightarrow 3$
(or $2\leftrightarrow 4$) relates the coefficients $a_1$ and $a_3$
to each other, we conclude that the correlator (\ref{1st}) is, in
principle, determined by {\it two a priori independent functions}
of the conformal cross-ratios.

This result is, of course, completely general, and must hold
perturbatively as well as non-perturbatively. In the following we
will study the radiative corrections to this correlator, and will
show that they are given in terms of a {\sl single} independent
coefficient function; the ratios of the ``quantum" parts of the
functions $a_1,a_3,b_2$ are completely fixed:
\begin{eqnarray}
  a_1^{\mbox{\scriptsize q. corr.}} &=& s{\cal F}(s,t)\;,
  \qquad a_3^{\mbox{\scriptsize q. corr.}} =t{\cal F}(s,t)\;,
  \qquad b_2^{\mbox{\scriptsize q. corr.}} =(1-s-t){\cal F}(s,t)
\nonumber
\end{eqnarray}
(see eq.(\ref{I.3}) below).
To this end we have to supplement the pure symmetry arguments
given above by some knowledge about the dependence of the
hypermultiplet correlator on the Grassmann variables
(G-analyticity), superconformal covariance and a new, dynamical
property, the so-called harmonic (H-)analyticity.

\setcounter{equation}0 \section{The hypermultiplet in harmonic
superspace}\label{HSS}

In the preceding section we showed that all the information about
the $N=4$ correlator (\ref{N4corre}) is contained in the $N=2$ one
(\ref{1st1}). There is another reason why we prefer to work with
$N=2$ rather than $N=4$ superfields. The non-renormalisation
theorem that we are going to prove in section \ref{HM} is based on
the insertion formula originating in the standard off-shell
Lagrangian formulation of field theory. The absence of an
off-shell formulation of $N=4$ SYM theory makes it difficult to
justify this procedure in $N=4$ superspace. However, there exists
an {\it off-shell} reformulation of the $N=4$ theory in terms of
$N=2$ harmonic superfields \cite{hh}.

We start by a brief review of the formulation of the $N=2$
hypermultiplet in harmonic superspace (the reader may wish to
consult \cite{bigbook} for the details). It can be described as a
superfield in the Grassmann (G-)analytic superspace \cite{hh} with
coordinates $x^{\alpha\dot\alpha}_A,\theta^{+\alpha},
\bar\theta^{+\dot\alpha},u^\pm_i$. Here $u^\pm_i$ are the harmonic
variables which form a matrix of $SU(2)$ and parametrise the
sphere $S^2\sim SU(2)/U(1)$. A harmonic function $f^{(p)}(u^\pm)$
of $U(1)$ charge $p$ is a function of $u^\pm_i$ which is invariant
under the action of the group $SU(2)$ (acting on the index $i$ of
$u^\pm_i$) and homogeneous of degree $p$ under the action of the
group $U(1)$ (acting on the index $\pm$ of $u^\pm_i$). Such
functions have infinite harmonic expansions on $S^2$ whose
coefficients are $SU(2)$ tensors (multispinors). The superspace is
called G-analytic since it only involves half of the Grassmann
variables, the $SU(2)$-covariant harmonic  projections
$\theta^{+\alpha} = u^{+i}\theta^{\alpha}_i,\
\bar\theta^{+\dot\alpha} = u^{+}_i\bar\theta^{\dot\alpha\; i}$.
In a way, this is the generalisation of the familiar concept of a
left- (or right-)handed chiral superfield depending on a different
half of the Grassmann variables, either the left-handed
($\theta_\alpha$) or right-handed ($\bar\theta^{\dot\alpha}$) one.

In this framework the hypermultiplet is described by a G-analytic
superfield of charge $+1$, $q^+(x_A,\theta^+,\bar\theta^+,u)$
(and its conjugate $\tilde q^+(x_A,\theta^+,\bar\theta^+,u)$
where $\tilde{}$ is a special conjugation on $S^2$ preserving
G-analyticity). G-analyticity can also be formulated as
differential constraints on the superfield:
\begin{equation}\label{Gan}
  D^+_\alpha\; q^+ = \bar D^+_{\dot\alpha}\; q^+ = 0
\end{equation}
where $D^+_\alpha = u^{+}_i D^{\alpha\; i}$, $\bar
D^+_{\dot\alpha} = u^{+i} D^{\dot\alpha}_i$. Note the similarity
between (\ref{Gan}) and the chirality condition (\ref{consW1}).
In fact, both of them are examples of what is called a ``short
multiplet" in the AdS/CFT language \cite{andfer}.

It is well-known that this $N=2$ supermultiplet cannot exist off
shell with a finite set of auxiliary fields \cite{nogo}. This only
becomes possible if an infinite number of auxiliary fields
(coming from the harmonic expansion on $S^2$) are present. On
shell these auxiliary fields are eliminated by the harmonic
(H-)analyticity condition (equation of motion)
\begin{equation}\label{EMo}
  D^{++}q^+ = 0\;.
\end{equation}
Here $D^{++}$ is the harmonic derivative on $S^2$ (the raising
operator of the group $SU(2)$ realised on the $U(1)$ charges,
$D^{++}u^+=0,\; D^{++}u^-=u^+$).

The reader can better understand the meaning of eq. (\ref{EMo})
by examining the general solution to the H-analyticity condition
on a (non-singular) harmonic function of charge $p$:
\begin{equation}\label{115}
  D^{++}f^{(p)}(u^\pm)=0 \ \Rightarrow \  \left\{
  \begin{array}{l}
    f^{(p)} = 0 \ \ \mbox{if}\ \ p<0\;; \\
    f^{(p)} = u^+_{i_1}\ldots u^+_{i_p}f^{i_1\ldots i_p} \ \ \mbox{if}\ \ p\geq
0\;.
  \end{array}
 \right.
\end{equation}
In other words, the solution only exists if the charge is
non-negative and it is a polynomial of degree $p$ in the harmonics
$u^+$. The coefficient $f^{i_1\ldots i_p}$ forms an irrep of
$SU(2)$ of isospin $p/2$. Thus, H-analyticity is just an $SU(2)$
irreducibility condition having the form of a differential
constraint on the harmonic functions.

Now it becomes clear why the combination of the G-analyticity
constraints (\ref{Gan}) with the H-analyticity one (\ref{EMo}) is
equivalent to the original on-shell hypermultiplet constraints
(\ref{hmdef}). Indeed, from (\ref{115}) one derives
(\ref{trivsol}) and then, by removing the arbitrary harmonic
commuting variables from both $q^+$ and $D^+, \bar D^+$, one
arrives at (\ref{hmdef}).

The crucial advantage of the harmonic superspace formulation is
that the equation of motion (\ref{EMo}) can be derived from an
{\it off-shell} action given by an integral over the G-analytic
superspace:
\begin{equation}\label{6.7.1}
S_{\mbox{\scriptsize HM}} = -\int
dud^4x_Ad^2\theta^+d^2\bar\theta^+\; \tilde q^{+}D^{++}q^+\;.
\end{equation}
This is the starting point for quantisation of the theory in a
straightforward way \cite{hsgr}. In particular, one can introduce
the following propagator (two-point function):
\begin{equation}\label{prop}
  \langle \tilde q^+_a(1)q^+_b(2)\rangle ={(12)\over
4\pi^2\; \hat x^{2}_{12}}\delta_{ab}\;.
\end{equation}
Here
$$
\hat x_{12} = x_{A1} - x_{A2} + {4i\over (12)}[(1^-2) \theta^+_1
\bar\theta^+_1 + (2^-1) \theta^+_2 \bar\theta^+_2 + \theta^+_1
\bar\theta^+_2 + \theta^+_2 \bar\theta^+_1]
$$
is a supersymmetric coordinate difference and, e.g., $(1^-2)=
u^{-i}_1 u^{+j}_{2}\epsilon_{ij}\ $. This propagator satisfies the
Green's function equation (suppressing the color indices),
\begin{equation}\label{Grf}
  D^{++}_1\langle \tilde q^+(1)q^+(2)\rangle = \delta^4(x_{12})\;
  \delta^2(\theta^+_1-\theta^+_2)\; \delta^2(\bar\theta^+_1-\bar\theta^+_2)\;
  \delta(u_1,u_2)\;.
\end{equation}
In deriving (\ref{Grf}) one makes use of the following property of
the harmonic derivative of a {\it singular} harmonic function:
$D^{++}_1\; (12)^{-1} = \delta(u_1,u_2)$ .

Let us now assume that the space-time points 1 and 2 are kept
apart, $x_1 \neq x_2$. Then the right-hand side of eq. (\ref{Grf})
vanishes and the two-point function (\ref{prop}) becomes
H-analytic:
\begin{equation}\label{HA2p}
   D^{++}_1\langle \tilde q^+(1)q^+(2)\rangle = 0 \quad \mbox{if } x_1 \neq x_2\;.
\end{equation}
This property can be extended to any correlation function
involving gauge-invariant composite operators made out of
hypermultiplets, ${\cal O}=\mbox{Tr} [(\tilde q^+)^p(q^+)^q]$:
\begin{equation}\label{HAnp}
  D^{++}_1\langle {\cal O}(1)\ldots\rangle = 0 \quad \mbox{if } x_1 \neq x_2,x_3,\ldots\;.
\end{equation}
In reality eq. (\ref{HAnp}) is a Schwinger-Dyson equation based on
the free filed equation (\ref{EMo}). Hence, its right-hand side
contains contact terms like in (\ref{Grf}), which vanish if the
space-time points are kept apart. So, H-analyticity is a {\it
dynamical} property of such $N=2$ correlators holding away from
the coincident points. It will play an important r{\^o}le in the
next section.

Finally, just a word about the other ingredient of the $N=4$
theory, the $N=2$ SYM multiplet. Here we do not need to go through
the details of how it is formulated in harmonic superspace and
subsequently quantised \cite{Zup}. We just recall that the
corresponding field strength is a (left-handed) chiral superfield
which is harmonic independent, $D^{++}W=0$. The $N=2$ SYM action
is given by the chiral superspace integral
\begin{equation}\label{SYMact}
  S_{\mbox{\scriptsize N=2 SYM}} =
{1\over 4\tau^2}\int d^4x_Ld^4\theta\;  {\rm Tr}\;W^2
\end{equation}
where $\tau$ is the (complex) gauge coupling constant.
\footnote{In fact, there exists an alternative form given by the
right-handed chiral integral $\int d^4x_Rd^4\bar\theta\; {\rm
Tr}\;\bar W^2$. In a topologically trivial background the
coupling constant becomes real and the two forms are equivalent
(up to a total derivative).}

\setcounter{equation}0 \section{ Superconformal covariance and
analyticity as the origin of non-renormalisation}\label{HM}

In this section we are going to derive the constraints on the
G-analytic correlator $G^{(N=2)}$ (\ref{1st1}) following from
superconformal covariance and H-analyticity. Applied to the lowest
component in its $\theta$ expansion (see (\ref{1st})),
H-analyticity simply means irreducibility under $SU(2)$, from
which one easily derives the three independent $SU(2)$ tensor
structures. Further, conformal covariance implies that their
coefficients are arbitrary functions of the conformal
cross-ratios (in the particular case (\ref{1st})
point-permutation symmetry reduces the number of independent
functions to two).

The combined requirements of superconformal covariance in
G-analytic superspace and H-analyticity have further, far less
obvious consequences which arise at the next level of the $\theta$
expansion of the correlator. They have been derived in
\cite{EHPSW} where it was found that one of the three coefficient
functions remains unconstrained while two linear combinations of
them have to satisfy certain first-order linear differential
constraints. The solution of the latter allows for some
functional freedom which cannot be fixed on general grounds. A
summary of these constraints as well as their explicit solution is
given in the Appendix.

Alternatively \cite{hssw}, one can apply a more efficient
procedure which, as it turns out, also completely fixes the above
freedom.
It is based on the insertion formula  \cite{ken1} (for an
explanation in the context of $N=2$ harmonic superspace see
\cite{hssw,ehssw3}). This formula relates the derivative of a
$4$-point correlator of the type (\ref{1st1}) with respect to the
(complex) coupling constant $\tau$ to a $4+1-$point correlator
obtained by inserting the $N=2$ SYM action (\ref{SYMact}):
\begin{equation}\label{6}
  {\partial\over\partial \tau}G^{(N=2)} \sim
  \int d^4x_{L0} d^4\theta_0\langle {\rm
Tr}\;W^2(0)|\widetilde q^+\widetilde q^+|q^+q^+|\widetilde q^+
\widetilde q^+ |q^+q^+\rangle\;.
\end{equation}
Recall that unlike the matter superfields $q^+$ which are
G-analytic and harmonic-dependent off shell, $W$ is chiral and
harmonic-independent. The integral in the insertion formula
(\ref{6}) goes over the chiral insertion point 0. As we shall see
later on, the combination of chirality with G-analyticity, in
addition to conformal supersymmetry and H-analyticity, impose
strong constraints on this five-point function.

Let us try to find out what could be the general form of a
$4+1$-point correlator $\Gamma^{(0|2,2,2,2)}$ which is chiral at
point 0 and G-analytic at points $1,\ldots,4$ (with $U(1)$
charges +2), has the corresponding superconformal properties and
is also H-analytic,
\begin{equation}\label{8}
   D^{++}_r\Gamma^{(0|2,2,2,2)} = 0\;, \quad \mbox{if } x_r\neq x_s\;,
   \quad  r,s=1,\ldots,4  \;.
\end{equation}
In particular, it should carry a certain R weight. Indeed, the
expansion of the matter superfield $q^+ = \phi^i(x)u^+_i +
\ldots$ starts with the physical doublet of scalars of the $N=2$
hypermultiplet which have no R weight. At the same time, the $N=2$
SYM field strength $W = \ldots + \theta\sigma^{\mu\nu}\theta
F_{\mu\nu}(x)$ contains the YM field strength $F_{\mu\nu}$ (R
weight 0) in a term with two left-handed $\theta$'s, so the R
weights of $W$ and of the Lagrangian equal 2 and 4, respectively.
{}From (\ref{6}) it is clear that this weight is compensated by
that of the chiral superspace measure $d^4x_Ld^4\theta$, so the
correlator on the left-hand side of eq. (\ref{6}) has the
expected weight 0.

The task now is to explicitly construct superconformal covariants
of R weight 4 out of the coordinates of chiral superspace
$x_{L0}\;,\ \theta^{i\alpha}_0$ at the insertion point 0 and of
G-analytic harmonic superspace $x_{A r}\;,\ \theta^{+\alpha}_r\;,\
\bar\theta^{+\dot\alpha}_r\;,\ u^\pm_{ri}\;$, $r=1,\ldots,4$ at
the matter points. To this end we need to know the transformation
properties of these coordinates under $Q$ and $S$ supersymmetry
(parameters $\epsilon$ and $\eta$, correspondingly). Since we are
only interested in the leading term in the $\theta$ expansion, it
is sufficient to examine the linearised transformations of the
Grassmann variables \cite{fradkin}:
\begin{eqnarray}
  \delta \theta^{i\alpha} &=& \epsilon^{i\alpha} + x^{\alpha\dot\beta}_L
\bar\eta^i_{\dot\beta} + O(\theta^2)\;,  \nonumber\\
  \delta\theta^{+\alpha} &=& u^+_i\epsilon^{i\alpha} +x^{\alpha\dot\beta}_A
  \bar\eta^i_{\dot\beta} u^+_i + O(\theta^2)\;,  \label{10}\\
 \delta\bar\theta^{+\dot\alpha} &=& u^+_i\bar\epsilon^{i\dot\alpha}
- u^{+}_i\eta^i_{\beta} x^{\beta\dot\alpha}_A  + O(\bar\theta^2)
\;. \nonumber
\end{eqnarray}

We remark that $Q$ supersymmetry acts as a simple shift, whereas
the $S$ supersymmetry terms shown in (\ref{10}) are shift-like.
Let us assume for the moment that we stay away from any
singularities in $x$-space (we shall come back to this important
point in a moment). Then we can use the four left-handed
parameters $\epsilon^{i\alpha}$ and $x^{\alpha\dot\beta}
\bar\eta^i_{\dot\beta}$ to shift away four of the six left-handed
spinors $\theta^{i\alpha}_0$ and $\theta^{+\alpha}_r$. This means
that our correlator effectively depends, to lowest order, on two
left-handed spinor coordinates. We can make this counting
argument more explicit by forming combinations of the $\theta$'s
which are invariant under $Q$ supersymmetry and under the
shift-like part of $S$ supersymmetry. $Q$ supersymmetry suggests
to use the differences
\begin{equation}\label{11}
  \theta^{\alpha}_{0r} = \theta^{i\alpha}_0u^+_{ri} - \theta^{+\alpha}_r\;,
\quad \delta_Q \theta^{\alpha}_{0r} = 0\;,  \quad r=1,\ldots,4\;.
\end{equation}
Then we can form the following two cyclic combinations of three
$\theta^{\alpha}_{0r}$:
\begin{equation}\label{12}
 (\xi_{12r})_{\dot\alpha} = (12)(x^{-1}_{0r}\theta_{0r})_{\dot\alpha} +
(2r)(x^{-1}_{01}\theta_{01})_{\dot\alpha} +
(r1)(x^{-1}_{02}\theta_{02})_{\dot\alpha}\;, \quad r=3,4
\end{equation}
where $x_{0r}\equiv x_{L0}-x_{Ar}$ are translation-invariant and
$(rs)\equiv u^{+i}_r u^{+j}_{s}\epsilon_{ij}$ are
$SU(2)$-invariant combinations of the space-time and harmonic
coordinates, correspondingly. It is now easy to check that
$\xi_{12r}$ are Q and S invariant to lowest order (i.e.,
shift-invariant):
\begin{equation}\label{13}
 \delta_{Q+S}\xi_{12r} = O(\theta^2)\;.
\end{equation}
Here one makes use of the harmonic cyclic identity
\begin{equation}\label{131}
  (rs)t_i + (st)r_i + (tr)s_i = 0\;.
\end{equation}

As a side remark we point out that the above counting of $Q$ and
$S$ shift-invariant Grassmann variables can also be applied to
the four-point function (\ref{1st1}). It depends on four
G-analytic $\theta^+$'s and their conjugates $\bar\theta^+$. This
number equals that of the $Q$ and $S$ supersymmetry parameters,
therefore we conclude that there exists no invariant combination
(under the assumption that we keep away from the coincident
space-time points). In other words, there are no nilpotent
superconformal G-analytic covariants having the properties of the
correlator (\ref{1st1}). This means that given the lowest
component (\ref{1st}) in the $\theta$ expansion of (\ref{1st1})
and using superconformal transformations we can uniquely
reconstruct the entire correlator (\ref{1st1}). A similar
argument applies to the $N=4$ correlator (\ref{N4corre}).

Let us now inspect the structure of the correlator
$\Gamma^{(0|2,2,2,2)}$ more closely. As we noted earlier, it has
R weight 4. In superspace the only objects carrying R weight are
the odd coordinates, so the $\theta$ expansion of our correlator
must start with the product of four left-handed $\theta$'s, i.e.,
the correlator should be nilpotent \cite{ehw,hssw}. Further,
superconformal covariance requires that the shift-like
transformations (\ref{10}) do not produce structures with less
than four $\theta$'s, so we must use all the four available
shift-invariant combinations (\ref{12}) (notice that they have R
weight 1, even though they are right-handed spinors). Thus, we
can write down the leading term in the correlator in the
following form:
\begin{equation}\label{14}
   \Gamma^{(0|2,2,2,2)} = (12)^{-2}\;{\xi_{123}^2 \xi_{124}^2   }\;
    F(x,u) + O(\theta^5\bar\theta)\;.
\end{equation}
The coefficient function $F$ depends on the space-time and
harmonic variables and carries vanishing $U(1)$ charges, due to
the explicit harmonic prefactor $(12)^{-2}$.

The nilpotent prefactor in (\ref{14}) can be expanded in terms of
$\theta_0$ and $\theta^+$. In fact, what contributes in the
insertion formula (\ref{6}) is just the purely chiral term in it:
\begin{eqnarray}
 (12)^{-2}\;{\xi_{123}^2 \xi_{124}^2   } &=& (\theta_0)^4\;
{R'\over
x^2_{01}x^2_{02}x^2_{03}x^2_{04}}
+ \,\,\mbox{terms containing $\theta^+$} \label{18}
\end{eqnarray}
where

\bear
R' &=& (12)^2(34)^2 x_{14}^2x_{23}^2 +(14)^2(23)^2x_{12}^2x_{34}^2
\non\\&&
+(12)(23)(34)(41)
\Bigl[x_{13}^2x_{24}^2-x_{12}^2x_{34}^2-x_{14}^2x_{23}^2\Bigr]
\non\\
&=& -\half\biggl\lbrace
(12)^2(34)^2
\Bigl[x_{12}^2x_{34}^2-x_{13}^2x_{24}^2-x_{14}^2x_{23}^2\Bigr]
\non\\&&
\hspace{21pt}
+ (13)^2(24)^2
\Bigl[x_{13}^2x_{24}^2-x_{12}^2x_{34}^2-x_{14}^2x_{23}^2\Bigr]
\non\\&&
\hspace{21pt}
+ (14)^2(23)^2
\Bigl[x_{14}^2x_{23}^2-x_{12}^2x_{34}^2-x_{13}^2x_{24}^2\Bigr]
\biggr\rbrace
\label{defRprime}
\ear\no
Here the polynomial $R'$ has been written in two different ways
using the harmonic cyclic identity (\ref{131}).
Note that the superconformal covariant (\ref{18}) is completely
permutation symmetric in the points $1,\ldots,4$, although this
is not obvious from the form of the left hand side.

Next, we substitute (\ref{18}), (\ref{defRprime}) in (\ref{6})
and carry out the integration over the insertion point. The
$\theta_0$ integral is trivial due to the Grassmann delta function
$(\theta_0)^4$ in (\ref{18}). The result is
\begin{equation}\label{21}
  {\partial\over\partial \tau}G^{(N=2)} \sim
  {\cal F}(s,t)\left[s\;{(12)^2(34)^2\over
x^4_{12}x^4_{34}} + t\;{(14)^2(23)^2\over x^4_{14}x^4_{23}}
 +(1-s-t)\;{(12)(23)(34)(41)\over
x^2_{12}x^2_{23}x^2_{34}x^2_{41}}\right]
\end{equation}
where
\begin{equation}\label{22}
  {\cal F}(s,t) = {1\over s}\;\int d^4x_0 {x^4_{12}x^4_{34}x^2_{14}x^2_{23}\over
x^2_{01}x^2_{02}x^2_{03}x^2_{04}}\; F(x,u)
\end{equation}
is an arbitrary conformally invariant four-point function (the
factor $1/s$ is introduced for convenience).

The reason why we have not indicated any harmonic dependence on
the left-hand side of (\ref{22}) has to do with our last
requirement, namely, the H-analyticity condition (\ref{8}) on the
four-point amplitude (\ref{21}):
\begin{equation}\label{19}
  D^{++}_r [(12)^2(34)^2{\cal F}] = D^{++}_r [(14)^2(23)^2{\cal F}] =
  D^{++}_r [(12)(23)(34)(41){\cal F}] = 0\;.
\end{equation}
Since the function ${\cal F}$ has vanishing $U(1)$ charges, from
(\ref{115}) one easily derives that it must be harmonic
independent, i.e., an $SU(2)$ singlet:
\begin{equation}\label{191}
  {\cal F} = {\cal F}(s,t)\;.
\end{equation}

In the analysis so far we have not taken into account possible
contact terms in the $4+1$ correlator $\Gamma^{(0|2,2,2,2)}$. As
pointed out in \cite{petske}, they may become important in the
context of the insertion formula (\ref{6}). Concerning the
G-analytic points 1,\ldots,4, we have decided to keep away from
the coincident points, and this assumption allowed us to impose
H-analyticity. Thus, we never see contact terms of the type
$\delta^4(x_{rs})$, $r,s=1,\ldots,4$. However, there can also be
contact terms involving the insertion point 0 and only one of the
matter points, i.e., singularities of the type $(\partial_0)^n
\delta^4(x_{0r})$. Since we are supposed to integrate over $x_0$
in the insertion formula (\ref{6}), such terms may result in a
non-contact contribution to the left-hand side. Therefore we must
investigate them in detail. \footnote{Contact terms of the
ultra-local type and their effect on the non-renormalisation of
two- and three-point functions have been studied in \cite{hssw}.}

In order to do this we have to adapt our construction of
nilpotent superconformal covariants. As before, we can profit from
the available $S-$supersymmetry freedom to shift away two of the
four $Q-$invariant combinations $\theta_{0r}$, but this time we
should be careful not to use singular (in space-time) coordinates
transformations. Let us suppose that we are dealing with a
contact term containing, e.g., $(\partial_0)^n \delta^4(x_{04})$.
In the vicinity of the singular point $x_0\sim x_4$ the matrix
$x_{04}$ is not invertible anymore, so we cannot shift away
$\theta_{04}$. However, since the four matter points are kept
apart, the matrices $x_{01}$, $x_{02}$ and $x_{03}$ still are
invertible. This allows us to shift away, e.g., $\theta_{01}$ and
$\theta_{02}\;$. Then, just as before, we will be left with only
two spinors, $\theta_{03}$ and $\theta_{04}$, all of which must
be used to construct the nilpotent covariant of R weight 4. Thus,
the latter is unique. This counting can also be done in a
manifestly $Q-$ and $S-$covariant way. We can still use the
shift-invariant combination $\xi_{123}$ from (\ref{12}) but not
$\xi_{124}$ because it is now singular. Then we replace the
latter by
\begin{equation}\label{192}
  \hat \theta_{04} = {x_{04} \over (12)} \xi_{124} = \theta_{04}
  +{(24)\over (12)}x_{04}x^{-1}_{01}\theta_{01} +
   {(41)\over (12)}x_{04}x^{-1}_{02}\theta_{02}
\end{equation}
which is regular at $x_{04}=0$. Thus, the contact analog of
(\ref{14}) is
\begin{equation}\label{193}
  \Gamma^{(0|2,2,2,2)}_{ \rm contact} = (\partial_0)^n
\delta^4(x_{04})\; \xi_{123}^2 \hat \theta_{04}^2\;
    F_{ \rm contact}(x,u) + O(\theta^5\bar\theta)
\end{equation}
(note that if there are no derivatives on the delta function, we
will only see the first term in (\ref{192}), but we need the
other two in the general case). Next, using the identity $\hat
\theta_{04}^2 = -x^2_{04} \xi^2_{124}/(12)^2$ and (\ref{18}) in
the insertion formula we obtain a contribution to $G^{(N=2)}$
having exactly the same form as (\ref{21}) but where now
\begin{equation}\label{221}
  {\cal F}_{ \rm contact}(s,t) = -{1\over s}\;\int d^4x_0 \; (\partial_0)^n
\delta^4(x_{04})\; {x^4_{12}x^4_{34}x^2_{14}x^2_{23}\over
x^2_{01}x^2_{02}x^2_{03}}\; F_{ \rm contact}(x,u)\;.
\end{equation}
As before, H-analyticity implies that this function must be
harmonic independent.

In the above construction of contact contributions we have singled 
out the matter point 4 but any other matter point can be obtained 
by cyclic permutations (it is obvious that two such terms cannot 
help each other to achieve superconformal covariance). Remarking 
that the harmonic polynomial $R'$ in (\ref{21}) is invariant under 
such permutations, we conclude that the possible contact terms 
{\it do not alter the form} of the four-point function (\ref{21}). 
\footnote{It is a separate issue whether contact terms of this 
type can actually arise in a supergraph calculation. So far 
experience at two and three loops \cite{hssw,ess} has shown that 
this is not the case. Nevertheless, the possibility remains open.} 

We remark that this argument also removes a possible loophole
which so far had remained in the proofs of the non-renormalisation
theorems for extremal and next-to-extremal four-point functions
as presented in \cite{ehssw3,ehsw}.

Thus the combination of superconformal
covariance, chirality, G- and H-analyticity within the context of
the insertion formula (\ref{6}) restricts the freedom in any
four-point correlator of the type $G^{(N=2)}$ to a single
function of the conformal cross-ratios. We recall that (\ref{6})
only gives the derivative of $G^{(N=2)}$ with respect to the
coupling. This means that from (\ref{21}) and (\ref{1st}) we can
read off the quantum corrections to the three coefficients:
\begin{equation}\label{I.3}
  a_1^{\mbox{\scriptsize q. corr.}} = s{\cal F}(s,t)\;,
  \qquad a_3^{\mbox{\scriptsize q. corr.}} =t{\cal F}(s,t)\;,
  \qquad b_2^{\mbox{\scriptsize q. corr.}} =(1-s-t){\cal F}(s,t)\;.
\end{equation}
Using the symmetry constraints (\ref{permut12}) we find that
${\cal F}$ should satisfy
\begin{equation}\label{symF}
  {\cal F}(s,t) = {\cal F}(t,s)={1\over t}{\cal F}({s\over t},{1\over
  t})\;,
\end{equation}
and we determine the remaining coefficients $a_2$, $b_1$ and
$b_3$. Finally, we express all the six coefficients in the $N=4$
correlator (\ref{sixcoeff}) in terms of a {\it single
unconstrained function} of the cross-ratios (with the symmetry
properties (\ref{symF})):
\begin{eqnarray}
  a_1&=& A_0 + s\;{\cal F}(s,t)\;, \nonumber\\
  a_2&=& A_0 + {\cal F}(s,t)\;, \nonumber\\
  a_3&=& A_0 + t\;{\cal F}(s,t)\;, \nonumber\\
  b_1&=& B_0 + (s-t-1)\;{\cal F}(s,t)\;, \nonumber\\
  b_2&=& B_0 + (1-s-t)\;{\cal F}(s,t)\;, \nonumber\\
  b_3&=& B_0 + (t-s-1)\;{\cal F}(s,t)\;, \label{23}
\end{eqnarray}
The constants correspond to the free-field part which cannot be
determined by the insertion formula. The direct tree-level
computation including disconnected diagrams yields
respectively:
\begin{eqnarray}
A_0 = \frac{4 (N_c^2 - 1)^2}{(2 \pi)^8}\;,\qquad B_0 = \frac{16
(N_c^2 - 1)}{(2 \pi)^8}\;. \label{contree}
\end{eqnarray}
The result for the connected correlator is obtained from this by
setting $A_0=0$.

Another way to state our non-renormalisation theorem is to say
that in (\ref{23}) one finds linear combination of the coefficient
functions which are non-zero at the free-field level, but vanish
for all radiative corrections.

\setcounter{equation}0 \section{The supergravity result and the
AdS/CFT correspondence} \label{Strong}

According to the AdS/CFT correspondence, the connected part of the
correlator of four stress-tensor multiplets in $N=4$ SYM in the
limit $N_c \rightarrow \infty$ and $\lambda = g^2 N_c$ large but
fixed should match a certain tree-level correlator in $N=8$ AdS
supergravity in five dimensions. More precisely, the leading
component of the CFT correlator (\ref{sixcoeff}) that we are
discussing should correspond to the correlator of four sets of
supergravity scalars lying in the ${\bf 20}$ of $SO(6)$. The
latter has recently been computed by Arutyunov and Frolov
\cite{AF}. The purpose of this section is to first rewrite the
result of  \cite{AF} in a much simpler form and then compare it
to eqs. (\ref{23}) from Section \ref{HM}. We find a perfect match
between the CFT prediction and the explicit AdS result which, in
our opinion, is new evidence for the validity of the
correspondence conjecture.

According to \cite{AF}, the two independent coefficients of
the (connected) amplitude are, e.g.,
\begin{equation}
\frac{a_1}{x_{12}^4 x_{34}^4} =
\frac{32 N_c^2}{2^8 \pi^{10}} \, \Bigl[-
\frac{1}{2} \frac{1}{x_{34}^2} \, D_{2211} + (\frac{t}{s} +
\frac{1}{s}) x_{12}^2 \, D_{3322} + \frac{3}{2} \, D_{2222} \Bigr]
\end{equation}
and
\begin{eqnarray}
\frac{b_2}{x_{12}^2 x_{34}^2 x_{14}^2 x_{23}^2} &=& \frac{32
N_c^2}{2^8 \pi^{10}} \, \biggl[ (\frac{t}{s} - \frac{1}{s} + \frac{s}{t}
- \frac{1}{t} - 6) \, D_{2222} + 4 x_{12}^2 \, D_{3322} + 4
x_{14}^2 \, D_{3223} \nonumber \\ &+& x_{12}^2 \, D_{2211}
(\frac{1}{x_{12}^2 \, x_{34}^2} - \frac{1}{x_{14}^2 \, x_{23}^2})
+ x_{14}^2 \, D_{2112} (\frac{1}{x_{14}^2 \, x_{23}^2} -
\frac{1}{x_{12}^2 \, x_{34}^2}) \nonumber \\ &+& x_{13}^2 \,
D_{2121} (\frac{1}{x_{12}^2 \, x_{34}^2} +
\frac{1}{x_{14}^2\,x_{23}^2}) \biggr]
\end{eqnarray}
and the four others can be obtained from these by symmetry
according to equations (\ref{permut12}). The scaling weights in
the subscript of the $D$-functions refer to points 1 through 4 in
the obvious order. These functions where first introduced in
\cite{dfmmr1}; they denote the basis integrals appearing in
tree-level four-point calculations in AdS supergravity. They can
be written in various forms, of which the most useful one for our
purpose is the following \cite{AFP}:

\begin{eqnarray}
D_{\Delta_1 \Delta_2 \Delta_3 \Delta_4} &=& K \int_0^\infty
{\rm d}t_1 \ldots {\rm d}t_4 t_1^{\Delta_1-1} \ldots
t_4^{\Delta_4 - 1}
S_t^{-{\frac{\Delta_1+\Delta_2+\Delta_3+\Delta_4}{2}}}
\exp{\Bigl[- {1\over S_t}\sum_{i<j} t_i t_j x_{ij}^2 \Bigr]}
\non\\
\label{Drep}
\end{eqnarray}
Here $S_t = t_1+t_2+t_3+t_4$, and the prefactor is (specialising
to the $4+1$ - dimensional case)
\begin{equation}
K = \frac{\pi^2 \Gamma(\frac{\Delta_1 + \, \ldots \,
+\Delta_4}{2} - 2)}{2 \Gamma(\Delta_1) \ldots \Gamma(\Delta_4)} \,
.
\end{equation}
Using this representation for $D_{1111}$ and integrating out the
global scaling variable $S_t$, one immediately obtains the Feynman
parameter representation of the standard one-loop box integral
with external ``momenta" $x_{12},x_{23},x_{34},x_{41}$. As is
well-known, this integral can be expressed in terms of logarithms
and dilogarithms \cite{thovel}; following \cite{ussdavplb298}, we
write the result as
\begin{equation}
\frac{1}{K_{1111}} \, D_{1111} \, = \, \frac{1}{x_{13}^2 \,
x_{24}^2} \, \Phi^{(1)}(s,t) \, ,
\end{equation}
where
\begin{equation}
\Phi^{(1)}(s,t) = {1\over \lambda} \Biggl\lbrace 2\Bigl({\rm
Li}_2(-\rho s) + {\rm Li}_2(-\rho t)\Bigr) +\ln {t\over s} \ln
{{1+\rho t}\over {1+\rho s}} + \ln (\rho s)\ln (\rho t) +
{\pi^2\over 3} \Biggr\rbrace \label{Phi1explicit}
\end{equation}
and \footnote{Here we assume $\lambda^2 > 0$; the case $\lambda^2
< 0$ requires an appropriate analytic continuation.}
\begin{equation}
\lambda(s,t) = \sqrt{(1-s-t)^2-4st}, \qquad \rho(s,t) = 2
\,(1-s-t+\lambda)^{-1} \label{deflambdarho} \, .
\end{equation}
${\rm Li}_2$ denotes the dilogarithm function.

Moreover, from the representation (\ref{Drep}) it is also obvious
that all the D-functions occurring in the formulas above for $a_1$
and $b_2$ can be obtained from ${D_{1111}\over K_{1111}}$ by
appropriate differentiations with respect to the $x_{ij}^2$ (this
fact was already noted in \cite{dfmmr1}). Applying this algorithm
we rewrite $a_1,b_2$ in term of (third-order) differential
operators acting on the basic function $\Phi^{(1)}(s,t)$. Next,
note that
\begin{equation}
\partial_s \, \Phi^{(1)}(s,t) \, =  \, \frac{1}{\lambda^2} \, \Bigl(
\Phi^{(1)}(s,t) \, (1 - s + t) + 2 \ln(s) - \ln(t) \, (s + t -
1)/s \Bigr) \label{sameop2}
\end{equation}
and similarly with $s \leftrightarrow t$. These identities are
sufficient to express inductively arbitrary derivatives of
$\Phi^{(1)}$ with respect to $s,t$ by $\Phi^{(1)}$ itself and
logarithmic terms. The final result of this procedure is
\begin{eqnarray}
a_1 &=& s \, {\cal F}(s,t) \, , \label{stronga} \\ b_2 &=&
\frac{16 N_c^2}{(2 \pi)^8} \,  + (1-s-t) \, {\cal F}(s,t)
\label{strongb}
\end{eqnarray}
where
\begin{eqnarray}
{\cal F}(s,t) &=& - \frac{16 N_c^2}{(2 \pi)^8 \, \lambda^6}\biggl\lbrace
  \Phi^{(1)}(s,t) \, 12 \, s \, t \, [ (1 + s + t) \lambda^2
+ 10 \, s \, t ] \nonumber \\
  &&\hspace{45pt} + \ln(s) \, 2 \, s \, [ (1 + t^2 - s - s \, t
+ 10 \, t) \lambda^2 + 30 \, s \, t (1 + t - s) ] \nonumber \\
  &&\hspace{45pt} + \ln(t) \, 2 \, t \, [ (1 + s^2 - t - s \, t
+ 10 \, s) \lambda^2 + 30 \, s \, t (1 + s - t) ] \nonumber \\
  &&\hspace{45pt} +[ (1 + s + t) \lambda^4 +
20 \, s \, t \, \lambda^2 ]\biggr\rbrace\ .
\end{eqnarray}

The supergravity result (\ref{stronga}), (\ref{strongb})
exactly matches the form of our general prediction (\ref{23}).
The first term on the right-hand side of (\ref{strongb}) agrees,
to leading order in $N_c$ \footnote{As it is not clear whether
the gauge group in the AdS/CFT correspondence is of unitary or
special unitary type, the supergravity calculation \cite{AF}
assumed $U(N_c)$ for simplicity.},
with the connected free-field part (\ref{contree});
the function ${\cal F}$, which is the sum of all
quantum corrections in the appropriate limit, has the required
permutation symmetry properties and occurs with the expected
simple prefactors.

\setcounter{equation}0 \section{Conclusions}

The correlator of four stress-tensor multiplets in $N=4$ SYM
originally contains six amplitudes which can be reduced to only
two independent functions quite trivially, just by exploiting the
obvious symmetries. On grounds of superconformal invariance,
Grassmann and harmonic analyticity we show that the quantum part
of the amplitudes is in fact universal to all of them. It is given
by one a priori arbitrary function of the conformal cross-ratios
which occurs in all six amplitudes with simple prefactors.

This property had been observed in weak coupling perturbation
theory at two and three loops and we verify it for strong coupling
by simplifying the supergravity result of \cite{AF}. We regard
this as new evidence in favour of the AdS/CFT conjecture.
It provides also a highly non-trivial check on the results of
\cite{AF}.

Alternatively, one can turn the argument around: The supergravity
result does obey constraints originating from harmonic
superspace. This makes us believe that there exists an
appropriate harmonic superspace formulation of AdS supergravity
in which all these properties become manifest.

We remark that $N=4$ supersymmetry has not played any particular
r\^ole here. What we actually used was $N=2$ superconformal
invariance, a property which a large class of $N=2$ finite
theories posses \cite{finite}. We would need $N=4$ conformal
supersymmetry if we wish to reconstruct the correlator of four
stress tensors (the top component in the $N=4$ SYM multiplet)
from that of the scalar composites considered here. If we
restrict ourselves to finite $N=2$ theories, we can still claim a
non-renormalisation property of the correlator of four
hypermultiplet bilinears.

We have carried out the analysis of contact terms in the context
of the reduction formula to the extent necessary for our
present purposes. By an explicit construction of the relevant
type of contact terms we have shown that, at least at the four-point
level, their appearance in the Intriligator formula would
not alter the form of the result, which remains proportional to
the universal polynomial $R'$. This also removes
the only possible loophole remaining in the
proof of the non-renormalisation theorems for extremal and
next-to-extremal four-point functions as presented in \cite{ehssw3,ehsw}.
Those contact terms could, however, in principle have
invalidated the two-and three-loop computations presented
in \cite{hssw,ess}. Since their results have been
confirmed by independent calculations \cite{gps,bkrs} we conclude
that, up to the three-loop level, contact terms of the
``malignant'' type are absent.

Finally, it would be of obvious interest to study whether
the universality property found here has non-trivial implications
for the operator product expansion.

\vspace{30pt}
\noindent
{\bf Acknowledgements:}\\
\noindent Burkhard Eden, Christian Schubert and Emery Sokatchev
are indebted to Paul Howe and Peter West for sharing their
knowledge on correlators of constrained superfields. BE and ES
profited form stimulating discussions with Dan Freedman and Eric
D'Hoker. Anastasios Petkou thanks the Alexander-von-Humboldt
Foundation for support.

\setcounter{equation}0 \section{Appendix: Differential constraints
on the four-point function}

The main result of this paper is that the four-point correlator
(\ref{N4corre}) is determined by a single function of the
space-time variables, apart from simple terms reflecting the
free-field part of the amplitudes. The argument we gave makes use
of the insertion formula (\ref{6}) and therefore explicitly
distinguishes free-field and quantum parts.

Alternatively \cite{EHPSW}, one can directly study the
consequences of superconformal covariance and H-analyticity on the
higher-level terms in the $\theta$ expansion of $N=2$ correlators
of the type (\ref{1st1}). The constraints found in this way in
\cite{EHPSW} are reproduced here and solved explicitly. As before,
we find a function $F(s,t)$ contributing to all six amplitudes in
the same way as shown in (\ref{23}), but there is additionally an
a priori arbitrary function of one variable, named $h$ in the
following. We express the free-field part of the amplitudes $a_1,
b_2$ in terms of these functions. The expression for $h$ found in
this way is not zero, whereby it is clear that it cannot be
omitted on general grounds. This, in our opinion, explains why
the direct approach to the four-point function inevitably yields
weaker constraints.

We start by introducing the following two linear combinations of
the coefficients $a_1,a_3,b_2$ of the $N=2$ correlator (\ref{1st})
\footnote{The cross-ratios used in \cite{EHPSW} are related to
(\ref{crrt}) as follows: $s' = t/s$, $t' = 1/s$.}:
\begin{equation}\label{A.chvar}
  \alpha = {1\over s}a_1 + {s-1\over t^2}a_3 + {1\over t}b_2\;,
  \qquad \beta = {1-s-t\over t^2}a_3 - {1\over t}b_2\;.
\end{equation}
As shown in \cite{EHPSW}, the full implementation of
H-analyticity combined with superconformal covariance leads to
the following constraints:
\begin{eqnarray}
  &&\beta_s = -s\alpha_s - t\alpha_t - \alpha\;, \label{A.6''}\\
  &&\beta_t = s\alpha_s + (t-1)\alpha_t + \alpha\;. \label{A.ctss}
\end{eqnarray}
These first-order coupled differential equations have an
integrability condition in the form of a second-order equation
for each function:
\begin{equation}\label{A.2nd'}
  \Delta \alpha = \Delta \beta = 0
\end{equation}
where
\begin{equation}\label{A.2nd}
  \Delta = s\partial_{ss} + t\partial_{tt} + (s+t-1)\partial_{st}
  + 2\partial_s + 2\partial_t\;.
\end{equation}

We substitute
\begin{equation}\label{A.4}
  \alpha(s,t) = {A(s,t)\over\lambda}\;, \quad \beta(s,t) =
  {B(s,t)\over\lambda}
\end{equation}
where $\lambda$ has been defined in (\ref{deflambdarho}). Then we
perform a transformation to the variables
\begin{eqnarray}
\xi &=& \frac{2\ s}{1-s-t+\lambda} \; , \nonumber\\ \eta &=&
\frac{2\ t}{1-s-t+\lambda} \label{strong2}
\end{eqnarray}
whose inverse is
\footnote{When inverting (\ref{strong2}) we have to
remember that $\lambda$ (\ref{deflambdarho}) is defined as a {\it
positive} square root. Correspondingly, the choice of sign in the
last of eqs. (\ref{A.7}), as well as the form of the first two,
determines the allowed domain of the variables $\xi,\eta$. The
use of such variables has been studied in
\cite{DT}.}
\begin{eqnarray}
  s&=&\frac{\xi }{(1+\xi )\ (1+\eta )}\;,  \nonumber\\
  t&=&\frac{\eta }{(1+\xi )\ (1+\eta )}\;,  \nonumber\\
  \lambda&=& \frac{1-\xi\eta }{(1+\xi )\ (1+\eta )} > 0 \;. \label{A.7}
\end{eqnarray}
This brings the first equation in (\ref{A.2nd'}) to its ``normal
form":
\begin{equation}\label{A.8}
   A_{\xi\eta}=0\;.
\end{equation}
Its general solution obviously is
\begin{equation}\label{A.9}
  A(\xi,\eta) = f(\xi) + g(\eta)
\end{equation}
where $f,g$ are arbitrary functions. So, we have found
\begin{equation}\label{A.9'}
  \alpha(\xi,\eta) = {1\over \lambda} [f(\xi) + g(\eta)]\;.
\end{equation}

Given $\alpha$, it is not difficult to solve for $\beta$ from the
first-order system (\ref{A.6''}), (\ref{A.ctss}) (the integration
constant can be absorbed into a redefinition of $f,g$). The result
is
\begin{equation}\label{A.9''}
  \beta(\xi,\eta) = -{1\over \lambda} \left[{\xi f(\xi)\over 1+\xi} +
  {g(\eta) \over 1+\eta}\right]\;.
\end{equation}

At this point we should recall that $a_1,a_3,b_2$ appearing in
(\ref{A.chvar}) in fact originate from the $N=4$ correlator
(\ref{sixcoeff}) and are therefore subject to the symmetry
requirements (\ref{permut12}). Under the permutations
(\ref{permut13}) we have
\begin{equation}\label{permut121}
  1 \  \rightarrow \ 3\ : \ \xi \  \rightarrow \ \eta
  \qquad \mbox{and}\qquad 1 \  \rightarrow \ 2\ :
  \ \xi \  \rightarrow \  -{1/(1+\eta)}\ , \ \eta \  \rightarrow
  \ -{(1+\xi)/\xi}\;.
\end{equation}
Taking this into account leads to identifying the two functions in
(\ref{A.9'}), (\ref{A.9''}):
\begin{equation}\label{A.00}
  {f(\xi)\over 3(1+\xi)}= -{\xi g(\xi)\over 3(1+\xi)}  \equiv  h(\xi)\;.
\end{equation}
Further, the two independent coefficients, e.g.,  $a_1$ and $b_2$
become
\begin{eqnarray}
  a_1&=& {\xi\over 1-\xi\eta}\left[ F(\xi,\eta)
  +2h(\xi) -h(\eta) + h\left(-{1/(1+\xi)}\right) +
   h\left(-{1/(1+\eta)}\right) \right]\;, \nonumber\\
  b_2&=& {1+\xi\eta\over 1-\xi\eta} \left[ F(\xi,\eta) +
  h\left(-{1/(1+\xi)}\right) + h\left(-{1/(1+\eta)}\right)\right]\nonumber\\
  &&- {1- 2\xi\eta\over 1-\xi\eta}\; [h(\xi) + h(\eta)]\;. \label{A.symabc}
\end{eqnarray}
Here the functions $F(\xi,\eta)$ and $h(\xi)$ satisfy the
following constraints:
\begin{eqnarray}
  && F(\xi,\eta)=F(\eta,\xi) = F\left(-{1/(1+\eta)},-{(1+\xi)/\xi}\right)\;, \nonumber\\
  && h(\xi)+ h\left(-{(1+\xi)/\xi}\right) + h\left(-{1/(1+\xi)}\right)
   = C_q -(A_0 + B_0)\;,\label{A.01}
\end{eqnarray}
where the constants on the r.h.s. of the second equation are the
free-field values from equation (\ref{23}), possibly plus a
quantum correction $C_q$. The functions $F,h$ may be split into a
free-field part
\begin{eqnarray}
h_0(\xi) &=& \frac{1}{3} \, (A_0 (\xi + \frac{1}{\xi}) - B_0) \nonumber \\
F_0(\xi,\eta) &=& \frac{A_0}{3} \, \frac{-\xi^3 + 3 \xi + 1}{\xi \, (\xi + 1)}
+ \frac{A_0}{3} \, \frac{-\eta^3 + 3 \eta + 1}{\eta \, (\eta + 1)} + B_0
\end{eqnarray}
and a quantum part $h_q, F_q\;$.

In conclusion, the direct approach to the four-point correlator
(\ref{1st1}) leaves more freedom than the one from Section
\ref{HM}. Indeed, comparing (\ref{A.symabc}) with (\ref{symF}),
(\ref{23}) we see that in the latter case ${\cal F}=F_q/\lambda$
and the {\it quantum part} of $h$ vanishes, which is our
non-renormalisation result. The fact that the argument given in
this Appendix applies to both the free-field and quantum part
explains why it cannot be as strong as the one based on the
insertion formula.

\end{document}